\documentclass[a4paper,onecolumn,11pt]{IEEEtran}
\usepackage{graphics}
\usepackage{graphicx}    
\usepackage{float}
\usepackage{cite}  

\usepackage{amsmath} 
\usepackage{amssymb} 
\usepackage{amsfonts}

\usepackage{color}

    \def\R{\ensuremath{\mathbb{ R}}}
    
    \def\I{\ensuremath{\mathbb{ I}}}

    \def\d{\ensuremath{\text{d}}}

    \newcommand{\ip}[2]{\left\langle #1 , #2 \right\rangle}
    
\begin{document}

\title{Balanced model order reduction for systems depending on a parameter{$^{1}$}}
\author{Carles~Batlle$^{2}$ \and
	N\'{e}stor~Roqueiro$^{3}$
	\thanks{$^{1}$ 	This paper is a preprint of a paper submitted to \textit{IET Control Theory and Applications}, and is subject to Institution of Engineering and Technology Copyright. If accepted, the copy of record will be available at IET Digital Library.}
	\thanks{$^{2}$Carles Batlle is with Departament de Matem\`{a}tiques, EPSEVG and IOC,  Universitat Polit\`{e}cnica de Catalunya --- BarcelonaTech, Vilanova i la Geltr\'{u},  Spain: \texttt{ carles.batlle@upc.edu, ORCID:0000-0002-6088-6187 }}		
	\thanks{$^{3}$ N\'{e}stor Roqueiro is with Departamento de Automa\c{c}\~{a}o e Sistemas, Universidade Federal de Santa Catarina, Florian\'{o}polis, Brasil: \texttt{ nestor.roqueiro@ufsc.br}}
}

\maketitle 

\begin{abstract}                
\noindent  We provide an analytical framework for balanced realization model order reduction of linear control systems which depend on an unknown parameter. Besides recovering known results for the first order corrections, we obtain explicit novel expressions for  the form of second order corrections for singular values and singular vectors.
The final result of our procedure is an order reduced model which incorporates the uncertain parameter. We apply our algorithm to the model order reduction of a linear system of  masses and springs with parameter dependent coefficients.
\end{abstract}

\section{INTRODUCTION}
\label{intro}
Order reduced models \cite{Schilders2008_MOR} are useful to simulate very large models using less computational resources, allowing, for instance, the exploration of parameter regions. 
The lower order model should have some desirable properties, such as being easily computable, preserving some of the structural properties of the full model and, more importantly, yielding an error with respect to the original model that can be bounded in terms of the complexity of the approximating model. 
In particular, for linear time-invariant MIMO systems, model order reduction (MOR) based on the truncation of balanced realizations preserves the stability, controllability and observability of the full model, and furthermore provides  bounds for the norm of the error system \cite{Antoulas2005SIAM}.

The computation of a balanced realization for a linear system relies on numerical linear algebra algorithms, and does not allow for  the presence of symbolic parameters in the model. Hence, if a system contains an uncertain parameter, appearing, for instance, due to a physical coefficient which is only known to belong to a given interval, or due to the specification of a working point in a nonlinear system, the balancing procedure must be carried out for each numerical value of the parameter. This results in a set of reduced order models, which are difficult to work with if they are to be used to design a controller and, in any case, the explicit dependence on the original parameter is lost in the reduced system. 

In this paper we work out an algorithm to obtain a reduced order model which incorporates the original, symbolical parameter through a polynomial of arbitrary degree. To this end, we solve each step of the balanced realization procedure in powers of the symbolical parameter, although for the last step, which involves a singular value decomposition (SVD), we only provide explicit expressions up to second order corrections. Up to our knowledge, the second order correction to the singular subspaces that we obtain has not been reported in the literature, and it may be useful in other applications of SVD.

The paper is organized as follows. Section \ref{balanced} reviews the steps of the computation of the balanced realization for linear systems, and how a reduced order model can be constructed from it. Section    \ref{secGramians} develops a power series expansion for each of the above steps. We give explicit algorithms for each step, except for the singular value decomposition, which we develop only to second order. Section \ref{example} applies the procedure to a system of masses and springs with parameter dependent coefficients, and, finally, we discuss our results and point to possible improvements in Section \ref{conclusions}.

\section{REVIEW OF THE BALANCED REALIZATION PROCEDURE}
\label{balanced}

Consider the nonlinear control system
\begin{eqnarray}
	\dot x &=& f(x)+g(x)u,\label{k1a}\\
	y&=&h(x),\label{k1b}
\end{eqnarray} 
with $x\in\R^N$, $u\in\R^M$, $y\in\R^P$ and $f(0)=0$.

The controllability function $L_c(x)$ is the solution of the optimal control problem
\begin{equation}
	L_c(x)=\inf_{u\in L^2((-\infty,0),\R^M)} \frac{1}{2} \int_{-\infty}^0 ||u(t)||^2\d t
	\label{pk1}
\end{equation}
subject to the boundary conditions $x(-\infty)=0$, $x(0)=x$ and the system  (\ref{k1a}). Roughly speaking, $L_c(x)$ measures the minimum 2-norm of the input signal necessary to bring the system to the state $x$ from the origin.  

As shown in \cite{Scherpen93}, $L_c$ obeys the Hamilton-Jacobi-Bellman PDE
\begin{equation}
	\partial_x L_c f + \frac{1}{2} \partial_x L_c g g^T \partial_x^T L_c = 0,\quad L_c(0)=0, \label{k2a}
\end{equation}
in a domain $U_c\subset\R^N$ which contains the origin and where the vector field $-(f+gg^T\partial_x^TL_c)$ is asymptotically stable. 

The observability function  $L_o(x)$ is the 2-norm of the output signal obtained when the system is relaxed from the state $x$
\begin{equation}
	L_o(x)=\frac{1}{2} \int_0^\infty ||y(t)||^2\d t = \frac{1}{2} \int_0^\infty ||h(x(t))||^2\d t,
	\label{pk2}
\end{equation}
with $x(0)=x$ and subjected to (\ref{k1a}) with $u=0$, that is, $\dot x = f(x)$. It obeys the Lyapunov PDE
\begin{equation}
	\partial_x L_o f + \frac{1}{2} h^T h = 0,\quad L_o(0)=0, \label{k2b}
\end{equation}
in a domain $U_o\subset\R^N$ around the origin where $f(x)$ is asymptotically stable. 

For linear control systems, 
\begin{eqnarray}
	\dot x &=& Ax+Bu,\label{pk3}\\
	y&=& Cx,\label{pk4}
\end{eqnarray}
assumed to be observable, controllable and Hurwitz, both $L_c(x)$ and $L_o(x)$ are quadratic functions
\begin{eqnarray}
	L_c(x)&=&\frac{1}{2}  x^T W_c^{-1} x,\label{pk5}\\
	L_0(x)&=&\frac{1}{2}  x^T W_o x,\label{pk6}
\end{eqnarray} 
where $W_c>0$ and $W_o>0$, the controllability and observability Gramians, are the solutions to the matrix Lyapunov equations
\begin{eqnarray}
	AW_c+W_cA^T +BB^T &=&0,\label{pk7}\\
	A^TW_o+W_oA+C^TC &=& 0.\label{pk8}
\end{eqnarray}

As shown by Moore (\cite{moore81}; see also \cite{LHP87}, \cite{VeKa1983} and \cite{Verriest2008}), the matrix $W_c$ provides information about the states that are easy to control (in the sense that signals $u$ of small norm can be used to reach them), while  $W_o$ allows to find the states that are easily observable (in the sense that they produce outputs of large norm). From the point of view of the input-output map given by (\ref{pk3}) (\ref{pk4}), one would like to select the states that score well on both counts, and this leads to the concept of \textit{balanced realization}, for which $W_c=W_o$.   

The balanced realization is obtained by means of a linear transformation $x=Tz$, with $T$ computed as follows:

\begin{enumerate}
	\item Solve the Lyapunov equations
	\begin{eqnarray}
		AW_c+W_cA^T +BB^T &=&0,\label{L1}\\
		A^TW_o+W_oA+C^TC &=& 0,\label{L2}
	\end{eqnarray}
	with solutions $W_c>0$, $W_o>0$.
	
	\item Perform  Cholesky factorizations of the Gramians:
	\begin{equation}
		W_c=XX^T,\quad W_o=YY^T.
		\label{balan1}
	\end{equation} 
	Notice that $X>0$ and $Y>0$.
	\item Compute the SVD of $Y^TX$:
	\begin{equation}
		Y^TX=U\Sigma V^T,
		\label{balan2}
	\end{equation}
	with $U$ and $V$ orthogonal and 
	\begin{equation}
		\Sigma=\text{diag}(\sigma_1,\sigma_2,\ldots,\sigma_N),\quad \text{with}\ \sigma_1>\sigma_2>\cdots > \sigma_N>0.
		\label{balan2b}
	\end{equation}
	The $\sigma_i$ are the \textit{Hankel singular values}, and their squares $\tau_i=\sigma_i^2$ are often referred to as the squared singular values of the control system.
	\item The balancing transformation is given then by
	\begin{equation}
		T=XV\Sigma^{-1/2},\quad \text{with}\ T^{-1}=\Sigma^{-1/2}U^TY^T.
		\label{balan3}
	\end{equation}
	\item The balanced realization is given by the linear system
	\begin{equation}
		\tilde A = T^{-1}AT,\quad \tilde B = T^{-1}B,\quad \tilde C=CT,
		\label{balan4}
	\end{equation}
	and in the new coordinates
	\begin{eqnarray}
		\tilde W_c &=& T^{-1}W_c T^{-T}=\Sigma,\label{balan5a}\\
		\tilde W_o &=& T^T W_o T = \Sigma.\label{balan5b}
	\end{eqnarray}
\end{enumerate} 
Notice that, in the balanced realization,
\begin{eqnarray}
	\tilde L_c(z) &=& \frac{1}{2}\sum_{i=1}^N \frac{z_i^2}{\sigma_i}=\frac{1}{2}z^T\Sigma^{-1}z,\label{balan6a}\\
	\tilde L_o(z) &=& \frac{1}{2}\sum_{i=1}^N \sigma_i z_i^2=\frac{1}{2}z^T \Sigma z,\label{balan6b}
\end{eqnarray}
so that the state with only nonzero coordinate $z_i$ is both easier to control and easier to observe than the state corresponding to $z_{i+1}$, for $i=1,2,\ldots,N-1$. If, for a given $r$, $1\leq r < N$, one has $\sigma_r\gg \sigma_{r+1}$, it may be sensible, from the point of view of the map between $u$ and $y$, to keep just the states corresponding to the coordinates $z_1,z_2,\ldots,z_r$, and this is what is known as balanced realization model order reduction.

$\mathcal{H}_\infty$-norm lower and upper error bounds of  the balanced truncation method are given by
\begin{equation}
	\sigma_{r+1} \leq  \|G(s)-G_r(s)\|_{\mathcal{H}_\infty} \leq  2 \sum_{i=r+1}^n \sigma_i, \label{classical bounds}
\end{equation}
where $\sigma_i, i=1,\dots, n$, are the Hankel singular values of the system \cite{Glover1984_IJC}\cite{Enns1984CDC} (see \cite{Antoulas2005SIAM} for a thorough review, and references therein). 
From these inequalities it follows that, in order to get the smallest error for the truncated system, one  should disregard the states associated with the smallest Hankel singular values
(but see \cite{MBF2014Automatica}\cite{OpRe2015TAC} for a tighter lower bound that sometimes might yield a better approximation).

If we denote by $\tilde A_r$ the upper-left square block of $\tilde A$ formed by the first $r$ rows and columns, and by $\tilde B_r$ and $\tilde C_r$ the matrices obtained from the first $r$ rows or columns of $\tilde B$ or $\tilde C$, respectively, the reduced system of order $r$ obtained by balanced truncation is given by
\begin{eqnarray}
	\dot Z_r &=& \tilde A_r Z_r + \tilde B_r u,\label{brs1}\\
	y &=& \tilde C_r Z_r,\label{brs2} 	
\end{eqnarray}
with $Z_r =(z_1,\ldots, z_r)$.

One of the  problems of the above procedure is that it does not allow for the presence of symbolic parameters in the problem, since the solution of the matrix equations involved relies on numerical methods. In this paper we address this issue, assuming that the linear system is given by matrices $A(m)$, $B(m)$ and $C(m)$ which depend analytically on the parameter $m$. 
This  may represent an uncertain physical coefficient (this is the case of the example in Section \ref{example}), or it may appear by considering an unspecified working point in the linearization of  a nonlinear system.
Indeed, assume that (\ref{k1a}) has a   curve of fixed points  $x=x(\xi)$, $u=u(x(\xi))$, with $\xi\in\R$ the parameter of the curve, \textit{i.e.}   such that
$$
f(x(\xi))+ g(x(\xi))u(x(\xi))=0, \quad \text{for all $\xi$ in an open set.}
$$ 
Consider now a given value $\xi_0$ of $\xi$, and let 
$z=x-x_0$, with $x_0=x(\xi_0)$, and $v=u-u(x(\xi_0))$. One  obtains immediately that the corresponding linearization of (\ref{k1a}) is given by
\begin{equation}
	\dot z = F(\xi_0) z + G(\xi_0)v,
	\label{linearized_system}
\end{equation}
where $F$, $G$ are, respectively, $N\times N$ and $N\times M$ matrices with elements

\begin{eqnarray}
	F_{ij}(\xi_0) &=& \partial_j f_i(x_0)+ \sum_{k=1}^M \partial_j g_{ik}(x_0)u_{0k}(x_0),\quad i,j=1,\ldots N,\label{linearA}\\
	G_{ij}(\xi_0) &=& g_{ij}(x_0), \quad i=1,\ldots N,\ j=1,\ldots M.\label{linearB}
\end{eqnarray}
Furthermore, writing $w=y-h(x_0)$, the linearization of (\ref{k1b}) yields
\begin{equation}
	w= H(\xi_0) z,\quad H_{ij}(\xi_0) = \partial_j h_i(x_0),\ i=1,\ldots, P,\ j=1,\ldots, N.
	\label{linearC}
\end{equation}

Let $\hat{\xi}_0$ be an specific, \textit{i.e.} numeric, value of $\xi_0$ that we take as a reference working point, and let $m=\xi_0-\hat{\xi}_0$, and define
$$
A(m)=F(\hat{\xi}_0+m),\   B(m)=G(\hat{\xi}_0+m),\   C(m)=H(\hat{\xi}_0+m).
$$

Our goal  is to develop a power series expansion  in $m$ of the balanced model order reduction algorithm for the linear input/output system given by $A(m)$, $B(m)$, $C(m)$.  This will facilitate the analysis of how much the important degrees of freedom vary when $m$ is changed and, more importantly, will yield a reduced order model, suitable for control design, which incorporates the dependence on $m$ in  an explicit way. A survey of other  approaches to this problem is presented in \cite{BeGuWi2015SIAM}.

\section{POWER SERIES EXPANSION FOR THE BALANCED REALIZATION}
\label{secGramians}
Following  the previous discussion,  consider the control system
\begin{eqnarray}
	\dot x &=& A(m) x + B(m) u,\label{eq1}\\
	y &=& C(m)  x ,\label{eq1C}
\end{eqnarray}
with $m$ a symbolic parameter.
The controllability Gramian will depend also on $m$, and will be given by the solution $W^c(m)$ to the Lyapunov equation
\begin{equation}\label{eq2}
	A(m)W^c(m)+W^c(m)A^T(m)+ B(m)B^T(m)=0.
\end{equation}
Assume that $A(m)$, $B(m)$ and $C(m)$ are analytic in $m$,
\begin{eqnarray}
	A(m) &=& \sum_{k=0}^\infty A_k m^k,\label{eq3}\\
	B(m)&=& \sum_{k=0}^\infty B_k m^k, \label{eq4}\\
	C(m)&=& \sum_{k=0}^\infty C_k m^k, \label{eq4C}
\end{eqnarray}
and let us look for likewise solutions of the form
\begin{equation}\label{eq5}
	W^c(m)=\sum_{k=0}^\infty W^c_k m^k.
\end{equation}
Using the formal identities
\begin{equation}\label{eq6}
	\sum_{k=0}^\infty a_k t^k \ \sum_{j=0}^\infty b_j t^j = \sum_{r=0}^\infty \left(\sum_{s=0}^r a_{r-s}b_s \right) t^r
	= \sum_{r=0}^\infty \left(\sum_{s=0}^r a_{s}b_{r-s} \right) t^r,
\end{equation}
and substituting the above expansions into (\ref{eq2}) one immediately obtains
\begin{equation}\label{eq7}
	\sum_{s=0}^r \left( A_{r-s} W^c_s + W^c_s A_{r-s}^T + B_{r-s}B_s^T  \right) =0, \quad \text{for $r=0,1,2,\ldots$}.
\end{equation}
These are equivalent to the set of Lyapunov equations
\begin{eqnarray}
	A_0 W^c_0 + W^c_0 A_0^T + B_0 B_0^T &=& 0,\label{eq8}\\
	A_0 W^c_r + W^c_r A_0^T + P_r &=& 0,\  r=1,2,\ldots,\label{eq9}
\end{eqnarray}
with
\begin{equation}\label{eq10}
	P_r = B_0 B_r^T + \sum_{s=0}^{r-1}\left(            
	A_{r-s} W^c_s + W^c_s A_{r-s}^T + B_{r-s} B_s^T
	\right),\quad r=1,2,\ldots
\end{equation}
These equations can be solved recursively to the desired order, starting with the zeroth order Lyapunov equation (\ref{eq8}). Observe that the internal dynamics is always given by $A_0$, and that it is only the effective control term $P_r$ the one that changes with the order.

Similarly, the observability Gramian $W^o(m)$ satisfies  
\begin{equation}\label{eq11}
	A^T(m)W^o(m)+W^o(m)A(m)+ C^T(m)C(m)=0,
\end{equation}
and its power series solution
\begin{equation}\label{eq12}
	W^o(m)=\sum_{k=0}^\infty W^o_k m^k,
\end{equation}
can be obtained recursively from
\begin{eqnarray}
	A^T_0 W^o_0 + W^o_0 A_0 + C^T_0 C_0 &=& 0,\label{eq8b}\\
	A^T_0 W^o_r + W^o_r A_0 + Q_r &=& 0,\  r=1,2,\ldots,\label{eq9b}
\end{eqnarray}
with
\begin{equation}\label{eq10b}
	Q_r = C^T_0 C_r + \sum_{s=0}^{r-1}\left(            
	A^T_{r-s} W^o_s + W^o_s A_{r-s} + C^T_{r-s} C_s
	\right),\quad r=1,2,\ldots
\end{equation}

After computing $W^c(m)$ and $W^o(m)$ at the desired order, the next step in the balancing transformation procedure is to compute their ``square roots", $X(m)$ and $Y(m)$, such that
\begin{eqnarray}
	W^c(m) &=& X(m)X^T(m),\label{choC}\\
	W^o(m) &=& Y(m)Y^T(m).\label{choO}
\end{eqnarray} 
If 
\begin{equation}\label{cho1}
	X(m)=\sum_{k=0}^\infty  X_k m^k,
\end{equation}
one gets
\begin{equation}\label{cho2}
	W^c_k = \sum_{s=0}^k X_{k-s}X^T_s,	
\end{equation}	
which, again, are solved recursively as
\begin{eqnarray}
	X_0 X_0^T &=& W_0^c,\label{cho3}\\
	X_k X_0^T + X_0 X_k^T &=& W_k^c - \sum_{s=1}^{k-1} X_{k-s}X_s^T.\label{cho4} 	
\end{eqnarray}
Similarly, for
\begin{equation}\label{cho5}
	Y(m)=\sum_{k=0}^\infty  Y_k m^k,
\end{equation}
one arrives at
\begin{eqnarray}
	Y_0 Y_0^T &=& W_0^o,\label{cho6}\\
	Y_k Y_0^T + Y_0 Y_k^T &=& W_k^o - \sum_{s=1}^{k-1} Y_{k-s}Y_s^T.\label{cho7} 	
\end{eqnarray}
Equations (\ref{cho3}) and (\ref{cho6}) are standard Cholesky equations, but (\ref{cho4}) and (\ref{cho7}) are not Lyapunov (or Sylvester) equations for $X_k$ or $Y_k$ because of the presence of $X_k^T$ and $Y_k^T$, respectively. 

Equations of the form $AX+X^TB=C$ for $X$ have been studied in \cite{VoIk2011}, where the problem is reduced to a sequence of low-order linear systems for the entries of $X$.
However, the conditions for the uniqueness of the solution stated in \cite{VoIk2011} are not satisfied by equations of the form of (\ref{cho4}). Indeed,  in order to solve (\ref{cho4})
one has to consider $\det (X_0+ \lambda (X_0^T)^T)=(1+\lambda)^N \det X_0$, which vanishes for $\lambda=-1$ and thus violates condition (2) of Theorem 3 in \cite{VoIk2011}. Notice, however, that the right-hand side of (\ref{cho4}) is a symmetric matrix. If one splits $X_k$ into symmetric, $S_k$, and skew-symmetric, $T_k$, parts, one gets, after some calculations, that they obey
\begin{eqnarray}
	S_k X_0^T  + X_0 S_k &=&  W_k^c - \sum_{s=1}^{k-1} X_{k-s}X_s^T,\label{cho10}\\
	T_k X_0^T + X_0 T_k  &=& 0.\label{cho11}
\end{eqnarray}
Equations (\ref{cho10}) and  (\ref{cho11}) are Lyapunov equations, and in fact the generic solution to (\ref{cho11}) is  $T_k=0$. Hence, we have that the solution to (\ref{cho4}) is given by
\begin{equation}
	X_k=S_k,
\end{equation}
with $S_k$ the solution to the Lyapunov equation (\ref{cho10}), and an analogous reasoning applies to the solution of  (\ref{cho7}).

The last nontrivial step in the balancing algorithm is the singular value decomposition (SVD) of the product $Y^T(m)X(m)$,
\begin{equation}\label{svd0}
	Y^T(m)X(m) = U(m) \Sigma(m) V^T(m),
\end{equation}
where
\begin{equation}\label{sing_values}
	\Sigma(m) = \text{diag}(\sigma_1(m) \geq \sigma_2(m)\geq \ldots \geq \sigma_n(m)>0),	
\end{equation}
and $U(m)$ and $V(m)$ are $N\times N$ orthogonal matrices, depending also on the parameter $m$. 


Let us denote by $R_k$ the coefficients of the power series of $Y^T(m)X(m)$,
\begin{equation}\label{cho8}
	Y^T(m)X(m) = \sum_{k=0}^\infty R_k m^k, 
\end{equation}
with 
\begin{equation}\label{cho9}
	R_k = \sum_{s=0}^k Y_{k-s}^T X_s = \sum_{s=0}^k Y_s^T X_{k-s}. 
\end{equation}
Let also
\begin{eqnarray}
	U(m) &=& \sum_{k=0}^\infty U_k m^k,\label{svd1}\\
	V(m) &=& \sum_{k=0}^\infty V_k m^k,\label{svd2}\\	
	\Sigma(m) &=& \sum_{k=0}^\infty \Sigma_k m^k.\label{svd3}	
\end{eqnarray}
Notice that the coefficients of the power series for $V^{-1}(m)$,
\begin{equation}\label{svd4}
	V^{-1}(m) = \sum_{k=0}^\infty \hat V_k m^k,	
\end{equation}
can be computed recursively from those of $V(m)$ as
\begin{eqnarray}
	\hat V_0 &=& V_0^{-1},\label{svd5}\\
	\hat V_k &=& - V_0^{-1}\sum_{s=1}^k V_s \hat V_{k-s},\quad k=1,2,\ldots\label{svd6}
\end{eqnarray}
provided that $V_0$ is invertible, which is the case since we are assuming that $V(m)$ is orthogonal for all $m$, and in particular for $m=0$. For $k=1$ and $k=2$ one has, explicitly,
\begin{eqnarray}
	\hat V_1 &=& - V_0^{-1} V_1 V_0^{-1},\label{svd6_1}\\
	\hat V_2 &=& - V_0^{-1} V_2 V_0^{-1} + V_0^{-1}V_1 V_0^{-1} V_1 V_0^{-1}.\label{svd6_2}	
\end{eqnarray}

However, we will not need to compute the coefficients of $V^{-1}(m)$, as we will presently see.
From now on we will consider approximations only up to second order. As it will be clear from our presentation, obtaining higher order approximations is immediate but involves expressions that become quite cumbersome.  
We will write
\begin{eqnarray}
	R(m) &=& R_0 + m R_1 + m^2 R_2,\label{svd10a}\\
	U(m) &=& U_0 + m U_1 + m^2 U_2,\label{svd10b}\\
	V(m) &=& V_0 + m V_1 + m^2 V_2,\label{svd10c}\\
	\Sigma(m) &=& \Sigma_0 + m \Sigma_1 + m^2 \Sigma_2,\label{svd10d}
\end{eqnarray} 
with the understanding that any higher order contribution is neglected.  From $R=U\Sigma V^T$ one gets the identities
\begin{eqnarray}
	RV &=&  U \Sigma,\label{svd11a}\\
	R^T U &=& V \Sigma,\label{svd11b}
\end{eqnarray}
which in turn inply
\begin{eqnarray}
	R^T R V &=& V \Sigma^2,\label{svd11c}\\
	R R^T U &=& U \Sigma^2. \label{svd11d}	
\end{eqnarray}
If we denote by $u_j^{(k)}$ the $j$th column vector of $U_k$, and by $v_j^{(k)}$ the one of $V_k$, equations (\ref{svd11a}) and (\ref{svd11b}) imply
\begin{eqnarray*}
	\lefteqn{	(R_0+ m R_1 + m^2 R_2)(v_j^{(0)}+ m v_j^{(1)}+m^2 v_j^{(2)})}\\ &=& (\sigma_j^{(0)}+m \sigma_j^{(1)}+ m^2 \sigma_j^{(2)})(u_j^{(0)}+m u_j^{(1)}+m^2 u_j^{(2)}),\\
	\lefteqn{	(R_0^T+ m R_1^T + m^2 R_2^T)(u_j^{(0)}+ m u_j^{(1)}+m^2 u_j^{(2)})}\\ &=& (\sigma_j^{(0)}+m \sigma_j^{(1)}+ m^2 \sigma_j^{(2)})(v_j^{(0)}+m v_j^{(1)}+m^2 v_j^{(2)}),	
\end{eqnarray*}
with $\sigma_j^{(k)}$ the $j$th element of the diagonal matrix $\Sigma_k$. At zeroth, first and second order in $m$ these equations boil down to
\begin{eqnarray}
	R_0 v_j^{(0)} &=& \sigma_j^{(0)} u_j^{(0)},\label{svd12a}\\
	R_0^T u_j^{(0)} &=& \sigma_j^{(0)} v_j^{(0)},\label{svd12b}\\
	R_1 v_j^{(0)}+ R_0 v_j^{(1)} &=& \sigma_j^{(0)} u_j^{(1)} + \sigma_j^{(1)} u_j^{(0)},\label{svd13a}\\
	R_1^T u_j^{(0)}+ R_0^T u_j^{(1)} &=& \sigma_j^{(0)} v_j^{(1)} + \sigma_j^{(1)} v_j^{(0)},\label{svd13b}\\
	R_0 v_j^{(2)}+R_1 v_j^{(1)}+R_2 v_j^{(0)} &=& \sigma_j^{(0)} u_j^{(2)}+ \sigma_j^{(1)} u_j^{(1)}+ \sigma_j^{(2)} u_j^{(0)},\label{svd14a}\\
	R_0^T u_j^{(2)}+R_1^T u_j^{(1)}+R_2^T u_j^{(0)} &=& \sigma_j^{(0)} v_j^{(2)}+ \sigma_j^{(1)} v_j^{(1)}+ \sigma_j^{(2)} v_j^{(0)}.\label{svd14b}
\end{eqnarray}
Furthermore, the orthogonality condition $U^T(m)U(m)=\I$ implies
\begin{eqnarray*} 
	U_0^T U_0 &=& \I,\\
	U_1^T U_0 + U_0^T U_1 &=& 0,\\
	U_2^T U_0 + U_1^T U_1 + U_0^T U_2 &=& 0,
\end{eqnarray*}
which, in terms of the column vectors, are
\begin{eqnarray}
	\ip{u_i^{(0)}}{u_j^{(0)}} &=& \delta_{ij},\label{svd15}\\
	\ip{u_i^{(1)}}{u_j^{(0)}} + \ip{u_i^{(0)}}{u_j^{(1)}} &=& 0,\label{svd16}\\
	\ip{u_i^{(2)}}{u_j^{(0)}} + \ip{u_i^{(1)}}{u_j^{(1)}} + \ip{u_i^{(0)}}{u_j^{(2)}} &=& 0,\label{svd17}
\end{eqnarray} 
where $\ip{a}{b}=a^T b$ is the standard Euclidean inner product in $\R^n$. In particular, for $i=j$ one gets, besides $|| u_j^{(0)}||^2 =1$,
\begin{eqnarray}
	\ip{u_i^{(1)}}{u_i^{(0)}} &=& 0,\label{svd18}\\
	\ip{u_i^{(2)}}{u_i^{(0)}} &=& -\frac{1}{2} ||u_j^{(1)}||^2,\label{svd19}
\end{eqnarray}
and similarly for the $v_j^{(k)}$,
\begin{eqnarray}
	\ip{v_i^{(1)}}{v_i^{(0)}} &=& 0,\label{svd20}\\
	\ip{v_i^{(2)}}{v_i^{(0)}} &=& -\frac{1}{2} ||v_j^{(1)}||^2.\label{svd21}
\end{eqnarray}

The inner product of $u_i^{(0)}$ with (\ref{svd13a}) yields
\begin{eqnarray*}
	\lefteqn{\ip{u_i^{(0)}}{R_1 v_i^{(0)}} + \ip{u_i^{(0)}}{R_0 v_i^{(1)}} }\\ &=& \sigma_i^{(0)} \ip{u_i^{(0)}}{u_i^{(1)}} + \sigma_i^{(1)} \ip{u_i^{(0)}}{u_i^{(0)}}
	= \sigma_i^{(0)}\cdot 0 + \sigma_i^{(1)}\cdot 1,
\end{eqnarray*}
from which
\begin{eqnarray*}
	\sigma_i^{(1)} &=& 	\ip{u_i^{(0)}}{R_1 v_i^{(0)}} + \ip{u_i^{(0)}}{R_0 v_i^{(1)}} = \ip{u_i^{(0)}}{R_1 v_i^{(0)}} + \ip{R_0^T u_i^{(0)}}{ v_i^{(1)}}\\
	&=& \ip{u_i^{(0)}}{R_1 v_i^{(0)}} +  \sigma_i^{(0)}\ip{ v_i^{(0)}}{ v_i^{(1)}}\\ &=&  \ip{u_i^{(0)}}{R_1 v_i^{(0)}} +  \sigma_i^{(0)}\cdot 0 = \ip{u_i^{(0)}}{R_1 v_i^{(0)}}.	
\end{eqnarray*}
Hence, the first-order correction to the singular values is given by \cite{stewart1991}
\begin{equation}\label{svd22}
	\sigma_i^{(1)} = \ip{u_i^{(0)}}{R_1 v_i^{(0)}} = \ip{v_i^{(0)}}{R_1^T u_i^{(0)}},
\end{equation}
where the second form can also be obtained operating from (\ref{svd13b}). In order to complete the first order correction one needs to compute the corrections to the singular subspaces, \textit{i.e.} the vectors $u_i^{(1)}$ and $v_i^{(1)}$. To compute $u_i^{(1)}$, we act on (\ref{svd13b}) with $R_0$ and then use (\ref{svd13a}) to get rid of $v_i^{(1)}$:
$$
R_0 R_0^T u_i^{(1)} + R_0 R_1^T u_i^{(0)} =  \sigma_i^{(1)}R_0 v_i^{(0)} + \sigma_i^{(0)} \left(   -R_1 v_i^{(0)} + \sigma_i^{(1)}u_i^{(0)}+\sigma_i^{(0)} u_i^{(1)}
\right).
$$
One obtains thus 
\begin{eqnarray}
	\lefteqn{ \left(R_0 R_0^T - (\sigma_i^{(0)})^2\I  \right) u_i^{(1)} }\nonumber \\ &=& - R_0R_1^T u_i^{(0)} + \sigma_i^{(1)} R_0 v_i^{(0)} - \sigma_i^{(0)} R_1 v_i^{(0)} + \sigma_i^{(0)}\sigma_i^{(1)} u_i^{(0)}\nonumber\\
	&=& 2 \sigma_i^{(0)}\sigma_i^{(1)} u_i^{(0)} - R_0R_1^T u_i^{(0)}- \sigma_i^{(0)} R_1 v_i^{(0)}.\label{svd23}
\end{eqnarray}
This is a system of $N$ equations for the $N$ components of $u_i^{(1)}$, but the equations are not independent. Indeed, from (\ref{svd11d}) one has, to zeroth order,
\begin{equation}\label{svd24}
	R_0 R_0^T u_i^{(0)} = (\sigma_i^{(0)})^2 u_i^{(0)},
\end{equation}
so that $(\sigma_i^{(0)})^2$ is an eigenvalue of $R_0 R_0^T$ and $R_0 R_0^T - (\sigma_i^{(0)})^2\I$ is not invertible. Assuming that the eigenvalues are simple, one must find an extra equation in order to be able to obtain $u_i^{(1)}$, and this is provided by (\ref{svd18}). Denoting by $Q_i^{(1)}$ the vector in the right-hand side of (\ref{svd23}),
\begin{equation}\label{svd24b}
	Q_i^{(1)} = 2 \sigma_i^{(0)}\sigma_i^{(1)} u_i^{(0)} - R_0R_1^T u_i^{(0)}- \sigma_i^{(0)} R_1 v_i^{(0)},
\end{equation}
it turns out that each $u_i^{(1)}$ can be uniquely computed as the solution to the system
\begin{equation}\label{svd25}
	\left(
	\begin{array}{c}
		R_0 R_0^T - (\sigma_i^{(0)})^2\I \\
		(u_i^{(0)})^T
	\end{array}
	\right) u_i^{(1)} = \left(\begin{array}{c} Q_i^{(1)} \\ 0 \end{array}\right).
\end{equation}
An explicit form of the solution to (\ref{svd25}) for the more general case of non-square matrices is given in \cite{LLM2008}.
Similarly, for $v_i^{(1)}$ one has
\begin{equation}\label{svd26}
	\left(
	\begin{array}{c}
		R_0^T R_0 - (\sigma_i^{(0)})^2\I \\
		(v_i^{(0)})^T
	\end{array}
	\right) v_i^{(1)} = \left(\begin{array}{c} P_i^{(1)} \\ 0 \end{array}\right),
\end{equation}
with 
\begin{equation}\label{svd24c}
	P_i^{(1)} = 2 \sigma_i^{(0)}\sigma_i^{(1)} v_i^{(0)} - R_0^TR_1 v_i^{(0)}- \sigma_i^{(0)} R_1^T u_i^{(0)}.
\end{equation}

Under the assumption that the singular values $\sigma_i^{(0)}$  are non-degenerate, \textit{i.e.} the solution spaces of equations (\ref{svd11c}) and (\ref{svd11d}) are one-dimensional, the above systems have unique solutions that can be numerically computed. Let us assume, for instance, that there is a vector $u\neq 0$ such that
$$
\left(
\begin{array}{c}
R_0 R_0^T - (\sigma_i^{(0)})^2\I \\
(u_i^{(0)})^T
\end{array}
\right) u = 0.
$$ 
This implies, in particular, that 
$$
(R_0 R_0^T - (\sigma_i^{(0)})^2\I)u=0,
$$
and hence, due to the non-degeneracy, $u=\lambda u_i^{(0)}$ for some $\lambda$, which contradicts the last relation $(u_i^{(0)})^T u=0$.

In order to obtain the second order corrections one has to work with (\ref{svd14a}), (\ref{svd14b}) and (\ref{svd19}). For instance, multiplying (\ref{svd14a}) with $u_i^{(0)}$, using (\ref{svd19}) and (\ref{svd18}), and taking into account that
$$
\ip{u_i^{(0)}}{R_0 v_i^{(2)}} = \ip{R_0^T u_i^{(0)}}{v_i^{(2)}} = \sigma_i^{(0)} \ip{v_i^{(0)}}{v_i^{(2)}} = -\frac{1}{2} \sigma_i^{(0)} ||v_i^{(1)}||^2,
$$ 
one gets the second order correction to the singular values of $R$ 
\begin{equation}\label{svd27}
	\sigma_i^{(2)} = \frac{1}{2}\sigma_i^{(0)}\left(  ||u_i^{(1)}||^2 -||v_i^{(1)}||^2\right) + \ip{u_i^{(0)}}{R_1 v_i^{(1)}+ R_2 v_i^{(0)}}.
\end{equation}
Notice that the right-hand side depends only on data from the zeroth and first order approximations, plus the second order perturbation $R_2$. One can obtain an equivalent expression, 
changing everywhere
$R_i \to R_i^T$ and $u_i^{(k)}\leftrightarrow v_i^{(k)}$, if one starts instead with (\ref{svd14b}), although the equality of both expressions, in contrast to the first order computation, is not obvious.

In order to compute the second order correction to the singular subspaces one must solve (\ref{svd14a}) and (\ref{svd14b}) for $u_i^{(2)}$ and $v_i^{(2)}$. Using the same techniques as in the first order computation one  obtains, for instance, that
$$
\left(R_0 R_0^T - (\sigma_i^{(0)})^2\I  \right) u_i^{(2)} = Q_i^{2}
$$
with
\begin{eqnarray}
	Q_i^{(2)} &=& - R_0 R_1^T u_i^{(1)} - R_0 R_2^T u_i^{(0)} +  \sigma_i^{(0)}\sigma_i^{(1)} u_i^{(1)}\nonumber\\ &+& \sigma_i^{(1)} R_0 v_i^{(1)} + 2 \sigma_i^{(0)}\sigma_i^{(2)} u_i^{(0)} 
	- \sigma_i^{(0)} R_1 v_i^{(1)} - \sigma_i^{(0)} R_2 v_i^{(0)}. \label{svd28}
\end{eqnarray}
Again, the equations are not independent and one must add condition (\ref{svd19}) to them. Under the same nondegeneracy conditions as for the first order correction, the $u_i^{(2)}$ are then the unique solution to
\begin{equation}\label{svd29}
	\left(
	\begin{array}{c}
		R_0 R_0^T - (\sigma_i^{(0)})^2\I \\
		(u_i^{(0)})^T
	\end{array}
	\right) u_i^{(2)} = \left(\begin{array}{c} Q_i^{(2)} \\ -\frac{1}{2}||u_i^{(1)}||^2 \end{array}\right).
\end{equation}
Similarly, the $v_i^{(2)}$ are given by the solution to
\begin{equation}\label{svd30}
	\left(
	\begin{array}{c}
		R_0^T R_0 - (\sigma_i^{(0)})^2\I \\
		(v_i^{(0)})^T
	\end{array}
	\right) v_i^{(2)} = \left(\begin{array}{c} P_i^{(2)} \\ -\frac{1}{2}||v_i^{(1)}||^2 \end{array}\right),
\end{equation}
with
\begin{eqnarray}
	P_i^{(2)} &=& - R_0^T R_1 v_i^{(1)} - R_0^T R_2 v_i^{(0)} + \sigma_i^{(0)}\sigma_i^{(1)} v_i^{(1)}\nonumber\\ &+& \sigma_i^{(1)} R_0^T u_i^{(1)} + 2 \sigma_i^{(0)}\sigma_i^{(2)} v_i^{(0)} 
	- \sigma_i^{(0)} R_1^T u_i^{(1)} - \sigma_i^{(0)} R_2^T u_i^{(0)}. \label{svd31}
\end{eqnarray}
Notice that the matrices appearing on the left hand-sides of (\ref{svd29}) and (\ref{svd30}) are the same than the ones in (\ref{svd25}) and (\ref{svd26}), respectively, and hence the solutions are unique.  

This procedure can be repeated to obtain higher order corrections in $m$. At order $m$, one obtains first an explicit expression for the corrections $\sigma_i^{(m)}$ to the singular values, and then one can write  systems of equations
for the corrections $u_i^{(m)}$ and $v_i^{(m)}$ to the singular vectors, with the same matrices appearing in previous orders but with different right-hand sides.

The final step of the procedure for the construction of the balanced realization is to use (\ref{balan3}) with (\ref{svd10a})---(\ref{svd10d}), keeping terms up to order $m^2$. Since the matrix
$\Sigma(m)$ is diagonal, $\Sigma(m)^{-1/2}$ is defined diagonal-wise, and for each entry $\sigma_i(m)$ we have, up to order $m^2$,
\begin{eqnarray}
	(\sigma_i(m))^{-1/2} &=& (\sigma_i^{(0)}+m \sigma_i^{(1)}+m^2 \sigma_i^{(2)})^{-1/2}	\nonumber \\
	&=& \frac{1}{(\sigma_i^{(0)})^{1/2}} - m \frac{ \sigma_i^{(1)}}{2 (\sigma_i^{(0)})^{3/2}} \nonumber\\ &+& m^2\left(
	- \frac{\sigma_i^{(2)}}{2 (\sigma_i^{(0)})^{3/2}}  
	+  \frac{3 (\sigma_i^{(1)})^2}{8 (\sigma_i^{(0)})^{5/2}}  	\right) + O(m^3)\label{svd32}\\
	&\equiv& s_i^{(0)}+ m s_i^{(1)}+ m^2 s_i^{(2)} + O(m^3). \label{svd33} 
\end{eqnarray}
Hence, 
\begin{equation}\label{svd34}
	\Sigma(m)^{-1/2}= S_0 + m S_1 + m^2 S_2,	
\end{equation}
with
\begin{equation}\label{svd34b}
	S_a = \text{diag} (s_i^{(a)}),\quad a=0,1,2.
\end{equation}

Up to order $m^2$, the matrix $T(m)$ for the transformation from the original $x$ coordinates to the balanced ones $z$, $x=Tz$, and its inverse $T^{-1}(m)$, are given by
$T(m)=T_2(m)+O(m^3)$ and $T^{-1}(m)= T_2^{-1}(m)+O(m^3)$, with
\begin{eqnarray}
	T_2(m) &=& X_0V_0S_0 + m (X_0V_0S_1+X_0V_1S_0+X_1V_0S_0)\nonumber\\
	&+& m^2 (X_0V_0 S_2+X_2V_0S_0+X_0V_2S_0 \nonumber\\ & & +X_0V_1S_1+X_1V_0S_1+X_1V_1S_0)\label{svd35T} 	\\
	&\equiv& T_0 + m T_1  + m^2 T_2,\label{svd36T}\\
	T_2^{-1}(m) &=& S_0 U_0^T Y_0^T + m (S_0 U_1^T Y_0^T + S_0 U_0^T Y_1^T + S_1 U_0^T Y_0^T)\nonumber\\
	&+& m^2 (S_0 U_0^T Y_2^T + S_0 U_2^T Y_0^T + S_2 U_0^T Y_0^T\nonumber\\ & & + S_1 U_1^T Y_0^T + S_1 U_0^T Y_1^T + S_0 U_1^T Y_1^T)\label{svd35iT} \\
	&\equiv & \hat T_0 + m \hat T_1 + m^2 \hat T_2,\label{svd36iT}
\end{eqnarray}	   

From these, the  approximation of the balanced realization, up to the second order in $m$,  is given (see (\ref{balan4})) by
\begin{eqnarray} 
	\tilde A_2(m) &=& \hat T_0 A_0 T_0 + m (\hat T_0 A_1 T_0 + \hat T_0 A_0 T_1 + \hat T_1 A_0 T_0)\nonumber\\
	&+& m^2 (\hat T_0 A_0 T_2 + \hat T_0 A_2 T_0 + \hat T_2 A_0 T_0\nonumber\\ & & + \hat T_0 A_1 T_1 + \hat T_1 A_0 T_1 + \hat T_1 A_1 T_0),\label{svd37A}\\
	\tilde B_2(m) &=&
	\hat T_0 B_0  + m (\hat T_0 B_1 + \hat T_1 B_0) + m^2 (\hat T_0 B_2 + \hat T_2 B_0 + \hat T_1 B_1),
	\label{svd37B}\\
	\tilde C_2(m) &=&
	C_0 T_0 + m (C_0 T_1 + C_1 T_0) + m^2 (C_0 T_2 + C_2 T_0 + C_1 T_1).
	\label{svd37C}
\end{eqnarray}

Matrices (\ref{svd37A})---(\ref{svd37C}) define a balanced realization of the original system which is exact for $m=0$ and approximate to order $m^2$ for $m\neq 0$. A reduced system of order $r$ is obtained by truncating this realization so that only the first $r$ states are conserved. For $m=0$ one has only the error  which comes from the truncation associated to the number of states, while for  $m\neq 0$ one has to add to this the errors introduced by the Taylor truncations in the  steps of the procedure.    

\section{APPLICATION: A SYSTEM OF MASSES AND SPRINGS}
\label{example}

We consider a system of $N$ masses $m_i$ and (linear)springs with constants $k_i$ and natural lengths $d_i$, so that the $i$th spring lies between masses $m_i$ and $m_{i+1}$,
$i=1,\ldots, N-1$, and the last spring connects mass $m_N$ to a fixed wall. We also add a linear dampers to each mass, with coefficients $\gamma_i$ and, furthermore, act with an external force $M$ on the first mass. The equations of motion are given by
\begin{eqnarray*}
	m_1 \ddot x_1 &=& -k_1 (x_1-x_2-d_1) - \gamma_1 \dot x_1 + F,\\
	m_2 \ddot x_2 &=& -k_2 (x_2-x_3-d_2)+ k_1 (x_1-x_2-d_1) - \gamma_2 \dot x_2,\\
	&\vdots&\\
	m_{N-1}\ddot x_{N-1} &=& - k_{N-1}(x_{N-1}-x_N - d_{N-1})+ k_{N-2}(x_{N-2}-x_{N-1} -d_{N-2})\\ & & - \gamma_{N-1}\dot x_{N-1},\\
	m_N \ddot x_N &=& -k_N (x_N-d_N)+k_{N-1}(x_{N-1}-x_N-d_{N-1})-\gamma_N \dot x_N.	
\end{eqnarray*}	 
After redefining the coordinates to absorb the lengths $d_i$ and introducing the canonical momenta $p_i=\dot x_i/m_i$, the system can be put in the first order form
\begin{equation}
	\dot X = \left(\begin{array}{c|c} 0_{N\times N} & \text{diag}(1/m_1,\ldots,1/m_N)\\
		&  \\
		K_{N\times N} & -\text{diag}(\gamma_1/m_1,\ldots,\gamma_N/m_N)\end{array} \right) X + B F,	
\end{equation}	 
where $X=(x_1,\ldots,x_N,p_1,\ldots, p_N)^T$, 
$$
B=(\underbrace{0,\ldots,0}_N,1,0,\ldots,0)^T,
$$
and 
$$
K=\left(
\begin{array}{cccccccc}
-k_1 & k_1 & 0 & 0 &\cdots & 0 & 0 & 0 \\
k_1 & - (k_1+k_2) & k_2 & 0 &\cdots & 0 & 0 & 0 \\
0 & k_2 & - (k_2+k_3) & k_3 & \cdots & 0 & 0 & 0\\
\vdots & \vdots & \vdots & \vdots &  &\vdots & \vdots & \vdots \\
0 & 0 & 0 & 0 & \cdots & k_{N-2} & -(k_{N-2}+k_{N-1}) & k_{N-1}\\
0 & 0 & 0 & 0 & \cdots & 0 & k_{N-1} & -(k_{N-1}+k_N)
\end{array}
\right)
$$
If we measure the velocity of the first mass, we have the output $y=C Z$ with $C$ 
$$
C=\left(
\begin{array}{cccccccc}
1 & 0 & \cdots & 0 & 0 & 0 & \cdots & 0 
\end{array}
\right).
$$

In order to obtain a  test of our whole  algorithm, we consider the set of physical constants given by
\begin{eqnarray*}
	k_i &=& 100(i+1),\ i=1,\ldots,N,\\
	m_i &=& i(1+m),\ i=1,\ldots,N,\\
	\gamma_i &=& 1,\ i=1,\ldots,N,	
\end{eqnarray*}
with $m$ the  parameter of the Taylor expansion. We set $N=10$, which yields a  system with $20$ states, and consider reduced systems with four states. %
Our procedure, which we have implemented entirely in \texttt{Matlab}, yields the reduced system, parametrized  by $m$, given by
%

\begin{equation}
	\begin{split}
		A_4 &=\left(
		\begin{matrix}
			- 0.28\, m^2 + 0.255\, m - 0.218 & 0.504\, m^2 - 0.84\, m + 2.06 \\
			- 0.504\, m^2 + 0.84\, m - 2.06 &  - 0.0393\, m^2 + 0.0548\, m - 0.0799 \\
			0.198\, m^2 - 0.193\, m + 0.181 & 1.01\, m^2 - 1.05\, m + 1.07 \\ 
			0.648\, m^2 - 0.745\, m + 0.862 & 0.0653\, m^2 - 0.0808\, m + 0.103  
		\end{matrix}\right.
		\\
		&\qquad \qquad 
		\left.
		\begin{matrix}
			0.198\, m^2 - 0.193\, m + 0.181 &  - 0.648\, m^2 + 0.745\, m - 0.862\\ 
			- 1.01\, m^2 + 1.05\, m - 1.07 & 0.0653\, m^2 - 0.0808\, m + 0.103\\
			- 0.143\, m^2 + 0.149\, m - 0.155 & 1.39\, m^2 - 2.14\, m + 4.91\\
			- 1.39\, m^2 + 2.14\, m - 4.91 &  - 0.106\, m^2 + 0.119\, m - 0.134
		\end{matrix}\right),
	\end{split}
	\label{A4}
\end{equation}

\begin{equation}
	B_4=\left(\begin{array}{c}  - 0.0362\, m^2 + 0.0505\, m - 0.143\\ 3.95\cdot 10^{-4}\, m^2 + 0.00639\, m - 0.0813\\ 0.0135\, m^2 - 0.0239\, m + 0.102\\ 0.00731\, m^2 - 0.0167\, m + 0.0922 \end{array}\right),
	\label{B4}
\end{equation}
and
\begin{equation}
	\begin{split}
		C_4 &= 
		\left(\begin{matrix}  - 0.0362\, m^2 + 0.0505\, m - 0.143 &  - 3.95\cdot 10^{-4}\, m^2 - 0.00639\, m + 0.0813  \end{matrix}\right. \\
		&\qquad\qquad \left.\begin{matrix} 0.0135\, m^2 - 0.0239\, m + 0.102 &  - 0.00731\, m^2 + 0.0167\, m - 0.0922  \end{matrix}\right).
	\end{split}\label{C4}
\end{equation}

\begin{figure}
	\begin{center}
		\includegraphics[scale=1.6]{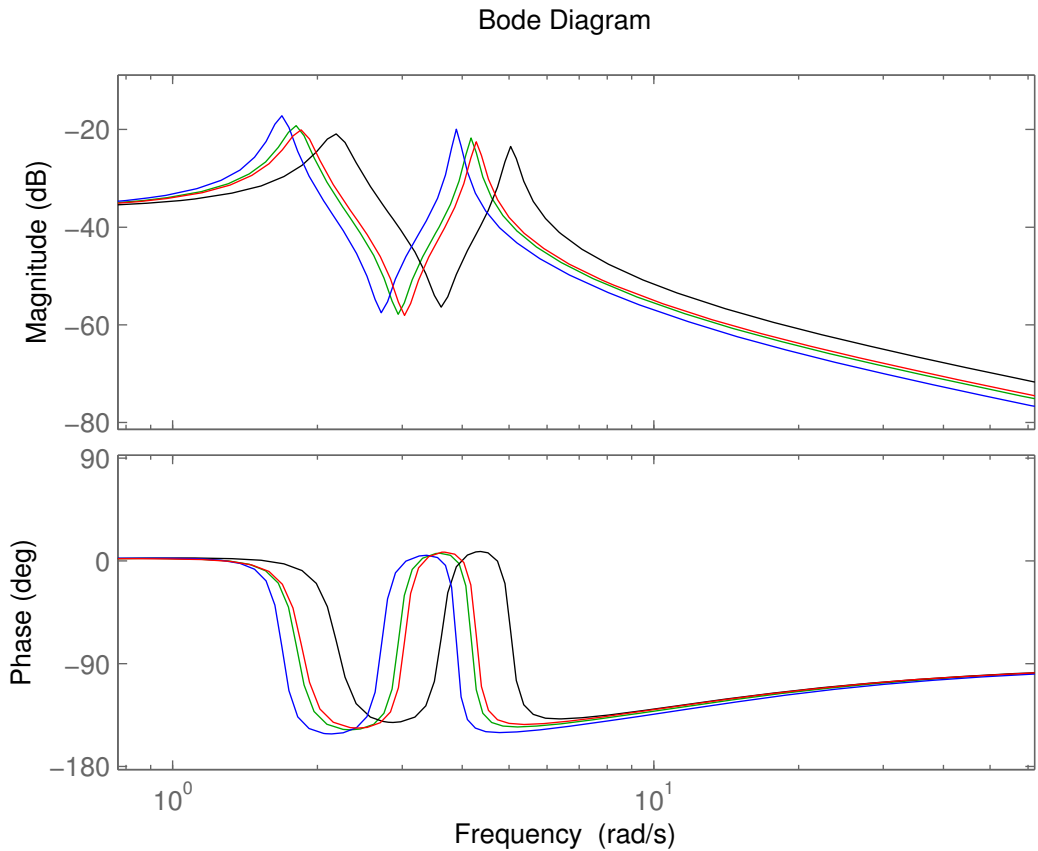}
	\end{center}
	\caption{Comparison of Bode plots for zeroth (black), first (blue) and second (red) order approximations for $m=0.5$, together with the \textit{exact} reduction of the system (green).}
	\label{fig1}
\end{figure}

Figure \ref{fig1} shows a detail of  the Bode diagrams for $m=0.5$ computed using the polynomial approximations of degree zero (black), one (blue) and two (red), together with the exact reduced system (green). It is clearly seen that the results improve as   the order of the polynomial approximation is increased. Notice that the zeroth order polynomial approximation is equivalent to considering $m=0$.

%

\section{CONCLUSIONS}
\label{conclusions}
We have developed a parameter dependent model order reduction algorithm based on the balanced realization approximation. The algorithm yields a reduced order model which can be used to design a controller valid for a range of values of the parameter. As a by-product, we have obtained an expression for the second order perturbation of the singular subspaces (see equations (\ref{svd29}) or (\ref{svd30})).  

We should point out that, from the point of view of simulating a large system, it may be better to compute the \textit{exact} reduced system for a given value of the parameter, since the truncation error of our second order polynomial approximation may become quite large for large $m$ (or even yield unstable reduced systems). Our procedure is thus more relevant for control design than for simulation.

Some trivial extensions of our work,  which we have not reported here for the sake of simplicity, include considering several parameters instead of one or computing some further higher order corrections of the parametrized SVD. 

We have not addressed the issue of the estimation of the error of the reduced model. Notice that this error involves both the truncation errors of the different steps of the algorithm and the error which comes from the truncation of the balanced realization. The latter is the only present for $m=0$, and is the one for which  bounds are well known. We currently do not know how to deal with the former, and how it could be integrated with the latter. However, the simulations of the system that we have presented, together with some simulations of the individual steps (not reported here) seem to indicate that the errors due to the different polynomial truncations go down when higher order approximations are used. We plan to address this issue by relating our construction to the general framework of \cite{BeDoGl1996} (see also \cite{Beck1996PhD}), and by comparing it to the approaches in \cite{BeGuWi2015SIAM}.

Our algorithm has an important limitation, namely that it can only be applied to stable systems. Application of coprime factorization techniques for parameter dependent systems \cite{PrempainCDC06}, which we plan to do in the future, could remove this drawback. 


\section*{Acknowledgements}
CB partially supported by  the Generalitat de Catalunya through project 2014 SGR 267 and by the Spanish government through DPI2015-69286-C3-2-R (MINECO/FEDER).
The authors would like to thank Yu. O. Vorontsov and Kh. D. Ikramov for making the Matlab code for their  ABST algorithm available to them.


\bibliography{biblio_MOR}          

\end{document}